\documentclass[%
 aip,
 amsmath,amssymb,
 reprint,%
]{revtex4-1}

\usepackage{graphicx}
\usepackage{dcolumn}
\usepackage{bm}
\usepackage{float}
\usepackage{hyperref}

\usepackage[utf8]{inputenc}
\usepackage[T1]{fontenc}
\usepackage{mathptmx}
\usepackage{etoolbox}

\makeatletter
\def\@email#1#2{%
 \endgroup
 \patchcmd{\titleblock@produce}
  {\frontmatter@RRAPformat}
  {\frontmatter@RRAPformat{\produce@RRAP{*#1\href{mailto:#2}{#2}}}\frontmatter@RRAPformat}
  {}{}
}%
\makeatother
\begin{document}

\preprint{AIP/123-QED}

\title[Chaos]{Linear and nonlinear causality in financial markets}
\author{Haochun Ma}
 \affiliation{Ludwig-Maximilians-Universit\"at, Department of Physics, Schellingstra{\ss}e 4, 80799 Munich, Germany}
  \affiliation{Allianz Global Investors, risklab, Seidlstra{\ss}e 24, 80335, Munich, Germany}

\author{Davide Prosperino}
 \affiliation{Ludwig-Maximilians-Universit\"at, Department of Physics, Schellingstra{\ss}e 4, 80799 Munich, Germany}
  \affiliation{Allianz Global Investors, risklab, Seidlstra{\ss}e 24, 80335, Munich, Germany}
  
\author{Alexander Haluszczynski}
  \affiliation{Allianz Global Investors, risklab, Seidlstra{\ss}e 24, 80335, Munich, Germany}

\author{Christoph Räth}
 \email{christoph.raeth@dlr.de}
 \affiliation{Deutsches Zentrum f{\"u}r Luft- und Raumfahrt (DLR), Institut f{\"u}r KI Sicherheit, Wilhelm-Runge-Stra{\ss}e 10, 89081 Ulm, Germany}

\begin{abstract}
Identifying and quantifying co-dependence between financial instruments is a key challenge for researchers and practitioners in the financial industry. Linear measures such as the Pearson correlation are still widely used today, although their  limited explanatory power is well known. In this paper we present a much more general framework for assessing co-dependencies by identifying and interpreting linear and nonlinear causalities in the complex system of financial markets.
To do so, we use two different causal inference methods, transfer entropy and convergent cross-mapping, and employ Fourier transform surrogates to separate their linear and nonlinear contributions. We find that stock indices in Germany and the U.S. exhibit a significant degree of nonlinear causality and that correlation, while a very good proxy for linear causality, disregards nonlinear effects and hence underestimates causality itself. The presented framework enables the measurement of nonlinear causality, the correlation-causality fallacy, and motivates how causality can be used for inferring market signals, pair trading, and risk management of portfolios. Our results suggest that linear and nonlinear causality can be used as early warning indicators of abnormal market behavior, allowing for better trading strategies and risk management.
\end{abstract}

\maketitle

\begin{quotation}
Within the complex system of financial markets, understanding the intricate ties between assets is crucial. Although the Pearson correlation has been a standard measure for these relationships, its linear approach might not fully represent the entire spectrum of causality. This study employs sophisticated causal inference algorithms and methods to differentiate between linear and nonlinear causal contributions. By examining major stock indices from Germany and the U.S., we uncover profound and possibly nonlinear linkages. More than presenting a new approach, this research indicates a significant shift in our perception and quantification of financial market behaviors. Such insights hold promise for refining market predictions, optimizing trading strategies, and improving portfolio risk management.
\end{quotation}

\section{\label{sec:introduction}Introduction}
\noindent The field of econophysics is garnering heightened attention in the physics domain, offering a novel lens to conventional financial methodologies \cite{jovanovic2018financial}. This emerging perspective draws from statistical physics tools, spanning signal processing, agent-based market frameworks, and random matrix theory \cite{wang2013random}. Understanding the co-dependence of financial assets is paramount across various finance sectors, especially when quantifying portfolio-associated risks \cite{mantegna1999introduction}. This development has seen industry practitioners keenly monitor the evolution of co-dependence metrics. Predominantly, mutual dependencies of financial instruments are characterized via the Pearson correlation of their return time series. However, there is increasing research underscoring the nonlinear characteristics of these series \cite{wang2018correlation}. Notably, \textcite{mantegna1995scaling} showed the power law scaling dynamics of financial indices' probability distributions, while \textcite{ghashghaie1996turbulent} pinpointed turbulent cascades in foreign exchange markets. Such insights challenge the adequacy of linear dependency metrics. Addressing this, \textcite{haluszczynski2017linear} segregated linear from nonlinear mutual information contributions using Fourier transform surrogates, aiming to quantify nonlinear correlations among financial assets. The authors demonstrated that the integration of nonlinear correlations into portfolio construction led to an increase in investment performance. 
A pressing query is the continued reliance on the Pearson correlation \cite{benesty2009pearson} as a causality proxy, given the intricate nature of causality measurement within dynamic systems. Granger's initial study in the 1960s \cite{granger2001essays} addressed the difference between causality and correlation, leading to the development of more advanced causal inference tools. This ranged from information-theoretic tools \cite{schreiber2000measuring} to state-space reconstruction models \cite{sugihara2012detecting}. While causal inference has mainly focused on determining causality \cite{ge2021dynamic}, the study of its linear versus nonlinear characteristics has not been performed in detail. Beginning work has been performed by \textcite{paluvs2007directionality} and \textcite{hlinka2014non}, who focused on mutual information to detect nonlinear dynamics in time series and evaluated nonlinearity contributions in climate connectivity.

In this paper, we analyze causality in financial markets by separating linear and nonlinear contributions to causality using Fourier transform surrogates. To do so, we use two different causal inference techniques and apply them to historical stock data of the German DAX and the U.S. Dow-Jones index. We also identify causality-based statistical properties of financial data and motivate how linear and nonlinear causality can be separated and measured. We find that while correlation is a good proxy for linear causality, nonlinear effects are disregarded and thus, significant amount of nonlinear causality is neglected. This is potentially dangerous when practitioners evaluate the risk of a portfolio only using correlation. Therefore, we propose a simple integration of causality measures into market signal inference, pair trading, and portfolio construction routines and show that they yield superior results.

\begin{figure*}
\centering
    \centering
    \includegraphics[width=\textwidth, keepaspectratio]{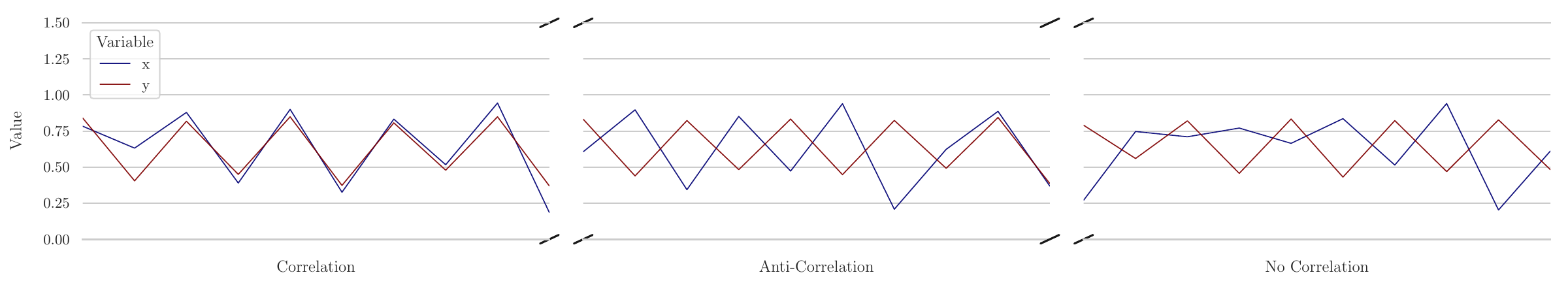}
    \centering
    \includegraphics[width=\textwidth, keepaspectratio]{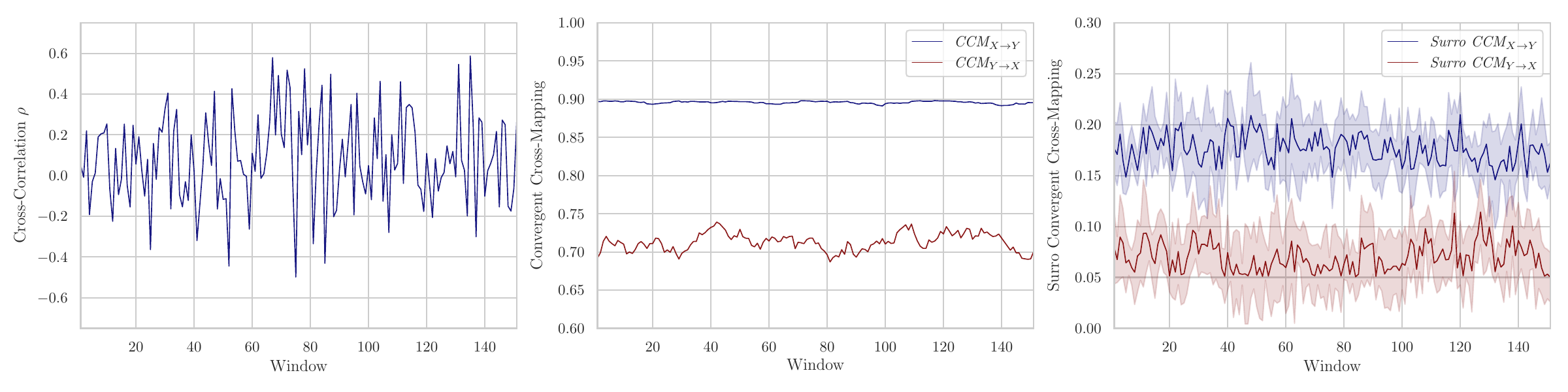}
\centering
\caption{Mirage correlations and causality. The top row shows different regimes of the coupled difference system defined in Equation \ref{eq:coupled_difference}. It appears that the variables are correlated in the first regime, anti-correlated in the second, and lose all coherence in the third. The bottom row shows the rolling correlation (left), causality (center), and linear causality (right). The causality is measured using Convergent Cross Mapping (CCM). While the correlation alternates between periods of positive, negative, and zero correlation, the causality in both directions stays stable over time. This also holds true for the linear causality. When comparing the measurements to the governing equations, we see that causality offers a more stable and accurate representation of the co-dependence between the two variables than correlation does.
}
\label{fig:coupled_difference}
\end{figure*}

\section{Methods}
\noindent We structure our methods section in three different parts: data, causality measures, linear and nonlinear decomposition, and financial frameworks.

\subsection{Data}
\noindent In this section, we describe the data used for this study. Before we apply our framework to real-world data, we demonstrate it on a synthetic example. Additionally, we use rolling windows in order to evaluate our analysis dynamically.

\subsubsection{Coupled Difference}
\noindent A simple example of a system that displays chaotic behavior is the coupled difference as introduced in \cite{lloyd1995coupled}. This system was also employed by \textcite{sugihara2012detecting} to illustrate \textit{Convergent Cross Mapping} (CCM), a causality inference method integral to this study. It is defined by the following two equations:

\begin{align}
\label{eq:coupled_difference}
\begin{split}
    x(t + 1) &= x(t) \cdot \left[r_x - r_x \cdot x(t) - \beta_{y \rightarrow x} \cdot y(t) \right]\\
    y(t + 1) &= y(t) \cdot \left[r_y - r_y \cdot x(t) - \beta_{x \rightarrow y} \cdot x(t) \right] \, ,
\end{split}
\end{align}

\noindent where the standard parameters are: $r_x = 3.8$, $r_y = 3.5$, $\beta_{y \rightarrow x} = 0.02$, and $\beta_{x \rightarrow y} = 0.1$. We selected this system due to its exhibition of so-called \textit{mirage correlations}, which means that variables may be positively coupled for long periods but can spontaneously become uncorrelated or decoupled. This can lead to problems when fitting models or inferring causality from observational data \cite{sugihara2012detecting}.

\begin{figure*}[!t]
     \centering
	\includegraphics[width=0.95\textwidth, keepaspectratio]{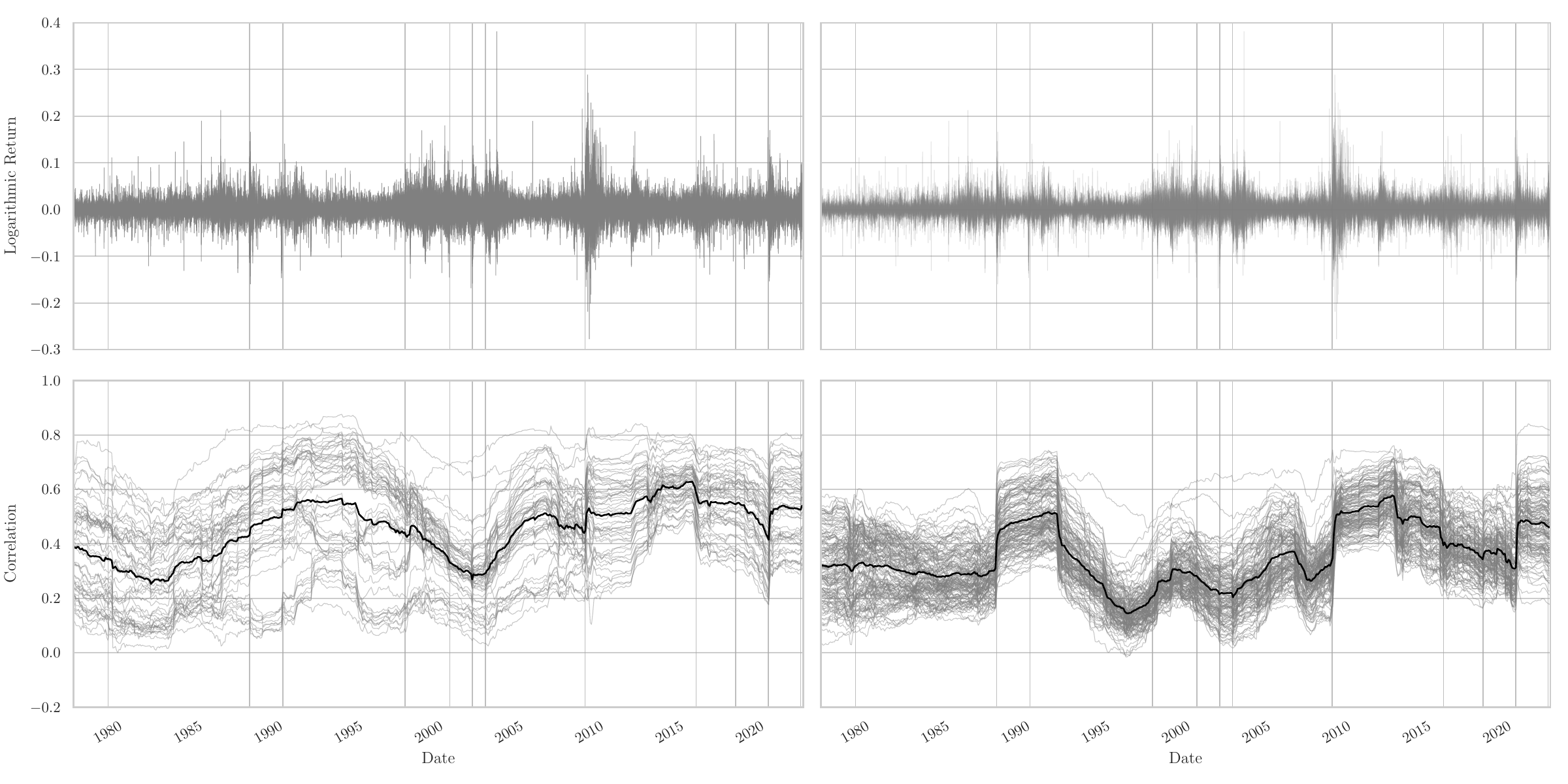}
    \caption{Historical Stock Returns and Correlation. The top row shows the logarithmic returns of the historical stock data of the German DAX (left) and the U.S. Dow-Jones (right) index, respectively. Each line represents the logarithmic return of one stock over time. The bottom row shows the pairwise correlations between the stocks. Each line represents the correlation between two stocks over time. The black line shows the average correlation inside the index. The vertical lines represent important economic or political events.}
    \label{fig:data_and_correlation}
\end{figure*}

\subsubsection{Financial Data}
\noindent For our real-world analysis, we select a subset of stocks from the DAX and Dow-Jones indices that represent the $30$ highest capitalized and thus most influential companies in Germany and the U.S., respectively. Beginning on January 19, 1973, our data consists of the daily closing prices of all stocks that were in the index through April 20, 2022, to provide a consistent universe of stocks over the entire period. This yields a total of $N_{DAX} = 11$ and $N_{DJ}=17$ time series with $12785$ data points. We would like to note that the survival bias \cite{brown1992survivorship} is negligible for our analysis. 

To ensure stationary time series, we convert the stock prices $p_t$ to logarithmic returns:

\begin{equation*}
\label{eq:log_returns}
    x_t =  \log{p_t} - \log{p_{t-1}} \, .
\end{equation*}

\noindent The time horizon of our data is long enough to examine a number of important market events---starting with the global recession of the early 1980s, it also includes Black Monday (October 19, 1987), when stock markets around the world collapsed for the first time since World War II. From 1997 to 2001, markets were characterized by excessive speculation and the overvaluation of many technology companies, which led to the \textit{dot-com bubble} \cite{delong2006short}. The bubble burst in 2002 with substantial price declines in July and September. Finally, our data includes the 2007/2008 subprime mortgage crisis, when the market declined from its all-time high in October 2007 and crashed after the collapse of Lehman Brothers on September 15, 2008. As a result of slowing growth of the GDP of China and the Greek debt default, investors sold shares globally between 2015 and 2016. The data further includes the so-called \textit{Volmageddon} on February 5, 2018, where a large sell-off in the U.S. stock market lead to a spike in implied market volatility \cite{krishnan2021vix}. Finally, the data includes the impact of the COVID-19 pandemic, which, among other events, triggered a sudden global stock market crash on February 20, 2020. In addition, our period under review also includes a number of important global political events. These include the fall of the Berlin Wall on November 9, 1989, which triggered the collapse of the Soviet Union, the attacks of September 11, 2001, and the Russian invasion of the Ukraine on February 24, 2022.

\subsubsection{Rolling Windows}
\noindent To obtain dynamically evolving results, we divide the data into overlapping rolling windows \cite{zivot2003rolling} and compute our measures for each interval following the approach by \textcite{haluszczynski2017linear}. We use a sliding window of $T_w = 1000$ trading days, which corresponds to roughly four years of data. The gap or stride between successive intervals is set to $\delta T = 20$ trading days, roughly amounting to a month. As such, the $w$-th interval is represented as:

\begin{equation}
\label{eq:rolling_windows}
    \mathbf{x}^{(w)} = \left(x_{1 + (w - 1)\cdot \delta T}, \dots, x_{T_w + (w - 1) \cdot \delta T}\right) \, ,
\end{equation}

\noindent which yields a total of $w=594$ overlapping windows. A (causality) measure $\psi\big(\mathbf{x}, \mathbf{y}\big) \mapsto \mathbb{R}$, which maps two time series to a real number, is thus transformed into a vector $\Psi \in \mathbb{R}^w$.

\begin{figure*}[!t]
	\includegraphics[width=0.95\textwidth, keepaspectratio]{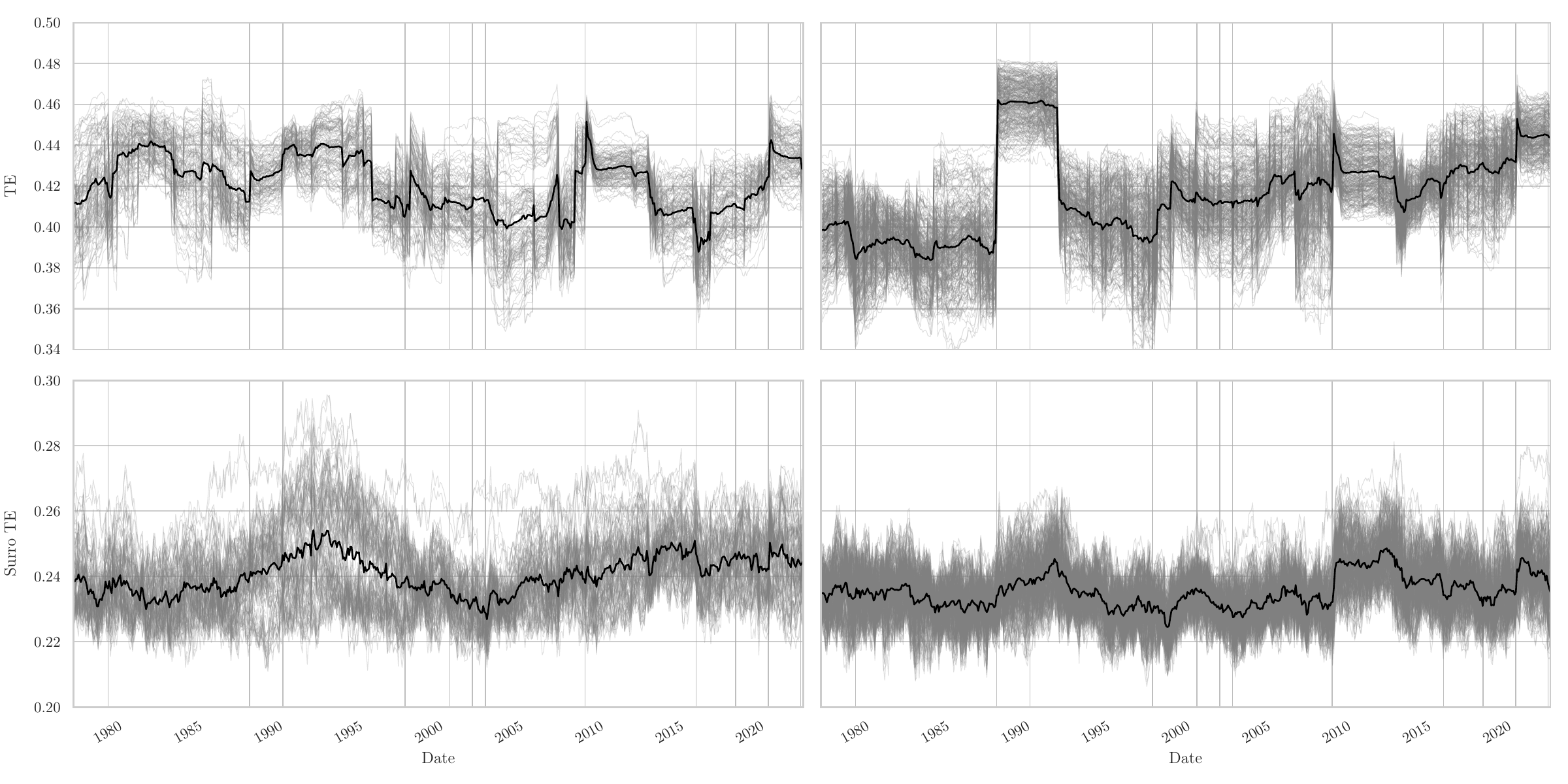}
    \caption{Transfer Entropy. The first row shows the historical TE of stocks within the German DAX (left) and the U.S. Dow-Jones (right) indices, respectively. Each line represents one direction of the TE between two stocks over time. The bottom row illustrates the corresponding surrogate TE. The vertical lines represent important economic or political events.
    }
    \label{fig:transfer_entropy}
\end{figure*}

\subsection{\label{sec:causality}Causality Measures}
\noindent We select two techniques that represent prominent categories currently used in causal inference \cite{runge2018causal}---however, it is important to note that our framework is applicable to any method capable of detecting nonlinear causality.

\subsubsection{Pearson Correlation}
\noindent Before describing the causal inference methods, we introduce the \textit{Pearson correlation} \cite{pearson1895vii}. We use it as a benchmark since it is still widely popular in the financial industry due to its simple calculation and interpretability \cite{benesty2009pearson}. It quantifies the strength and direction of the linear relationship between two variables. It is computed as follows:

\begin{equation}
\label{eq:correlation}
    \rho\left(\mathbf{x}, \mathbf{y}\right) \equiv \frac{\sum_{i=t}^T (x_t - \bar{x})(y_t - \bar{y})}{\sqrt{\sum_{t=1}^T (x_t - \bar{x})^2} \sqrt{\sum_{t=1}^T (y_t - \bar{y})^2}} \, ,
\end{equation}

\noindent where $x_t$ denotes the stock returns at time $t$ and $\bar{x} = \frac{1}{T} \sum_{t=1}^T x_t$ signifies their expected value. The correlation is normalized and bounded to the interval $[-1,1]$ and thus allows direct comparisons across pairwise correlations between different stocks. As shown by \textcite{bonett2000sample} a sample size of $T \leq 56$ is sufficient to estimate the measure reliably.

\subsubsection{Transfer Entropy}
\noindent \textit{Transfer Entropy} (TE) is a powerful information-theoretic measure introduced by \textcite{schreiber2000measuring} which has gained popularity in the field of causal inference, particularly in the analysis of time series data. TE provides a way to quantify the directed flow of information between variables, which allows assessing causal relationships in a probabilistic framework. The TE from $X$ to $Y$ is defined as:

\begin{equation*}
\textit{TE}_{X \rightarrow Y} = H(Y_{t+1}, Y_t) + H(Y_t, X_t) - H(Y_{t+1}, Y_t, X_t) - H(Y_t) \, ,
\end{equation*}

\noindent where $H(Y_{t+1}, Y_t)$, $H(Y_t, X_t)$, $H(Y_{t+1}, Y_t, X_t)$, and $H(Y_t)$ are the joint and marginal entropies of the respective variables. To facilitate comparison between different estimations of TE, we apply the subsequent normalization:

\begin{equation}
\textit{TE}_{X \rightarrow Y} = \frac{H(Y_{t+1}, Y_t) + H(Y_t, X_t) - H(Y_{t+1}, Y_t, X_t) - H(Y_t)}{\sqrt{H(Y_{t+1}, Y_t) \cdot H(X_{t+1}, X_t)}} \, .
\end{equation}

\noindent The normalization to $\left[0, 1\right]$ stems from our understanding of TE as an asymmetric causal measure. This interpretation aligns with the concept of covariance, which, when rescaled, results in the normalized form, the aforementioned Pearson correlation \cite{pearson1895vii}.

\begin{figure*}[!t]
	\includegraphics[width=0.95\textwidth, keepaspectratio]{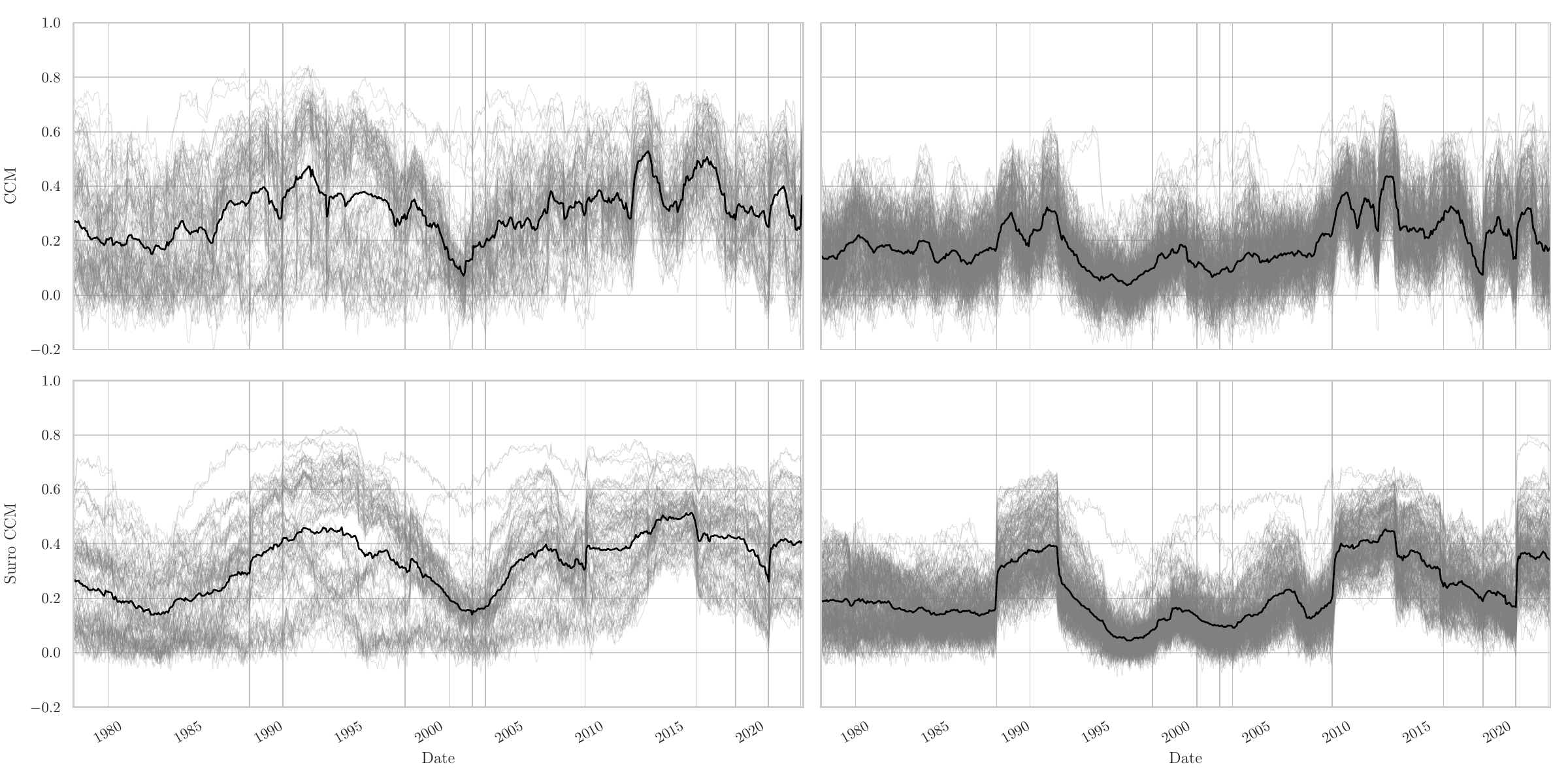}
    \caption{Convergent Cross Mapping. The setup of this figure is analogous to Fig. \ref{fig:transfer_entropy}.
    }
    \label{fig:convergent_cross_mapping}
\end{figure*}

We would like to point out that the calculation of empirical probability densities $p$ and hence information-theoretic measures raise unexpected difficulties exceeding the scope of this work. While it is common to use histograms with equally distributed bins to estimate densities, \citet{2021mynter} showed that this method potentially leads to biases since the estimation is dependent on the partition details---hence, finding a robust estimator is non-trivial. However, for the purpose of our research, we find that equally distributed bins perform reasonably well. Furthermore, it is worth mentioning that TE might capture false causalities depending on the dimension of conditioning \cite{paluvs2007directionality}.

\subsubsection{Convergent Cross Mapping}
\noindent \textit{Convergent Cross Mapping} (CCM) is an influential technique utilized for causal inference within the realm of complex dynamical systems \cite{sugihara2012detecting}. It aims to reveal causal connections between variables by reconstructing the dynamics that underlie them. CCM operates on the premise that variables with causal links will exhibit similar dynamical behavior, leading to a notion referred to as \textit{shadowing}. 

The underlying idea is based on Takens' theorem, which states that the entire state space can be reconstructed from a single embedded coordinate of the system, also called a \textit{shadow manifold} \cite{takens2006detecting}. Due to transitivity, two coordinates within a system can then be mapped to each other by neighboring states in their respective shadow manifolds---this allows for cross prediction. The quality of the prediction, evaluated using the Pearson correlation, quantifies the strength of the causal relationship. The algorithm of CCM can be outlined as follows:

\begin{enumerate}
\item \textbf{Time Delay Embedding} Embed the time series data of $X$ and $Y$ into higher-dimensional spaces using the embedding dimension $\kappa$ and time delay $\tau$.
    \item \textbf{Library Construction} Create a library of vectors from the reconstructed state space $\mathbf{X}$, denoted as $\mathcal{L}_{\mathbf{X}}$, and a library of vectors from the reconstructed state space $\mathbf{Y}$, denoted as $\mathcal{L}_{\mathbf{Y}}$.
    \item \textbf{Nearest Neighbor Selection} For each vector $\mathbf{X}(i)$ in the shadow manifold $\mathcal{M}_{\mathbf{X}}$, find its nearest neighbor in $\mathcal{M}_{\mathbf{Y}}$, denoted as $\mathbf{Y}(j)$. Similarly, for each vector $\mathbf{Y}(k)$ in $\mathcal{M}_{\mathbf{Y}}$, find its nearest neighbor in $\mathcal{M}_{\mathbf{X}}$, denoted as $\mathbf{X}(l)$.
    \item \textbf{Cross Mapping} Assess the predictability of $X$ based on $Y$ by comparing the distances between the vector pairs $\mathbf{X}(i)$ and $\mathbf{Y}(j)$, and the vector pairs $\mathbf{Y}(k)$ and $\mathbf{X}(l)$. A statistical measure, such as the correlation coefficient $\rho$, can be used to quantify the predictability.
    \item \textbf{Convergence Analysis} Repeat the cross mapping procedure for different library lengths. Evaluate the correlation as a function of the number of points used and assess the convergence of the results. The convergence of the cross mapping indicates the presence of a causal relationship between $X$ and $Y$.
\end{enumerate}

\noindent In the original application of CCM, convergence typically requires visual inspection. However, we've implemented a more systematic approach using expanding windows. For a given vector of correlations $\mathbf{\rho}$ of size $n$, we calculate the standard deviation within each window. Convergence is determined if the standard deviation consistently decreases, eventually falling below a predefined threshold $\theta$. If convergence is achieved, the mean of the last $s$ values is calculated to smooth any outliers. Conversely, if convergence is not reached, the causality measure in CCM is set to zero. This process is mathematically expressed as:

\begin{equation}
\label{eq:convergent_cross_mapping}
    \textit{CCM}_{X \rightarrow Y} \equiv 
    \begin{cases}
    \frac{1}{n} \sum_{i=1}^s \mathbf{\rho}_{n - s + i} \, &\text{if $\mathbf{\rho}$ converges}\\ 
    0 \, &\text{otherwise} 
    \end{cases}
    \in \left[-1, 1\right] \, .
\end{equation}

\noindent This process automates the evaluation of CCM causality for various connections within a system at a reasonable speed. To standardize the measure and render it comparable with other non-directional causal inference methods, the correlation distance, denoted as $d = \sqrt{2 \left(1 - \rho\right)}$, can be employed.

\noindent CCM's effectiveness in identifying causal relationships within time series data is affected by multiple aspects. The presence of noise or missing values in the data can alter the outcomes \cite{monster2017causal}, and the choice of appropriate embedding dimensions $\kappa$ and time delays $\tau$ is subject to the characteristics of the specific dataset \cite{wallot2018calculation}. For example, the optimal value for $\tau$ can be determined by finding the first local minimum in the \textit{Mutual Information} (MI) respective to $\tau$. Additionally, the \textit{False Nearest Neighbor} (FNN) algorithm can help finding the smallest embedding dimension that maintains the attractor's structure, ensuring that neighboring points in the original time series stay neighbors in the embedded version \cite{kennel1992determining}.

\begin{figure*}[!t]
	\includegraphics[width=0.95\textwidth, keepaspectratio]{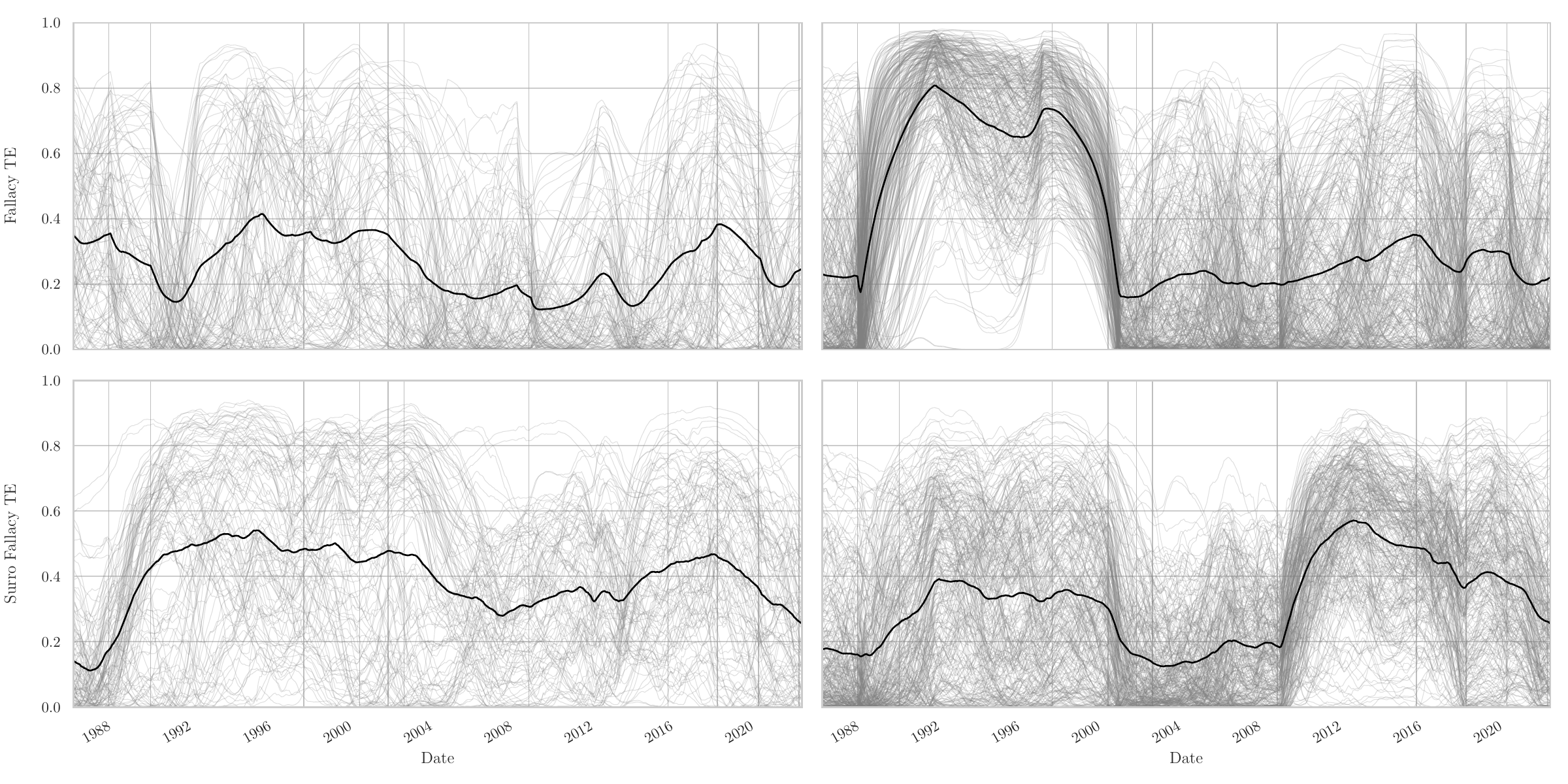}
    \caption{Fallacy Transfer Entropy. The first row shows the historical fallacy TE of stocks within the German DAX (left) and the U.S. Dow-Jones (right) indices, respectively. Each line represents one direction of the fallacy TE between two stocks over time. The bottom row illustrates the corresponding surrogate TE. The setup of this figure is analogous to Fig. \ref{fig:transfer_entropy}.
    }
    \label{fig:transfer_entropy_fallacy}
\end{figure*}

\subsubsection{Limits of Causality Measures}
\noindent We would like to emphasize our recognition of the limitations associated with the causal inference techniques we are presenting, as well as the broader challenges inherent in causal inference. Nonetheless, the purpose of this paper is to utilize these methods as a means to demonstrate a framework for dissecting causality into linear and nonlinear components within the context of finance. It is important to note that this paper does not delve into assessing the accuracy of these methods in capturing genuine causal relationships, nor does it explore the robustness of the methods themselves. Despite their drawbacks, these two methodologies have shown successful applications across various real-world scenarios \cite{mccracken2014convergent}. For in-depth analyses on TE and CCM, we recommend referring to \textcite{overbey2009effects} and \textcite{krishna2019inferring}, respectively.

Additionally, we acknowledge that TE and CCM operate with reconstructed spaces and have theoretical vulnerabilities when applied to variables within an attractor \cite{mccracken2014convergent}. Nevertheless, the analysis conducted in this paper relies on simulated data rather than a purely theoretical foundation. For a comprehensive discussion on the efficacy of state-space reconstruction methods in establishing causality, we direct interested readers to \textcite{cummins2015efficacy}.

\begin{figure*}[!t]
	\includegraphics[width=0.95\textwidth, keepaspectratio]{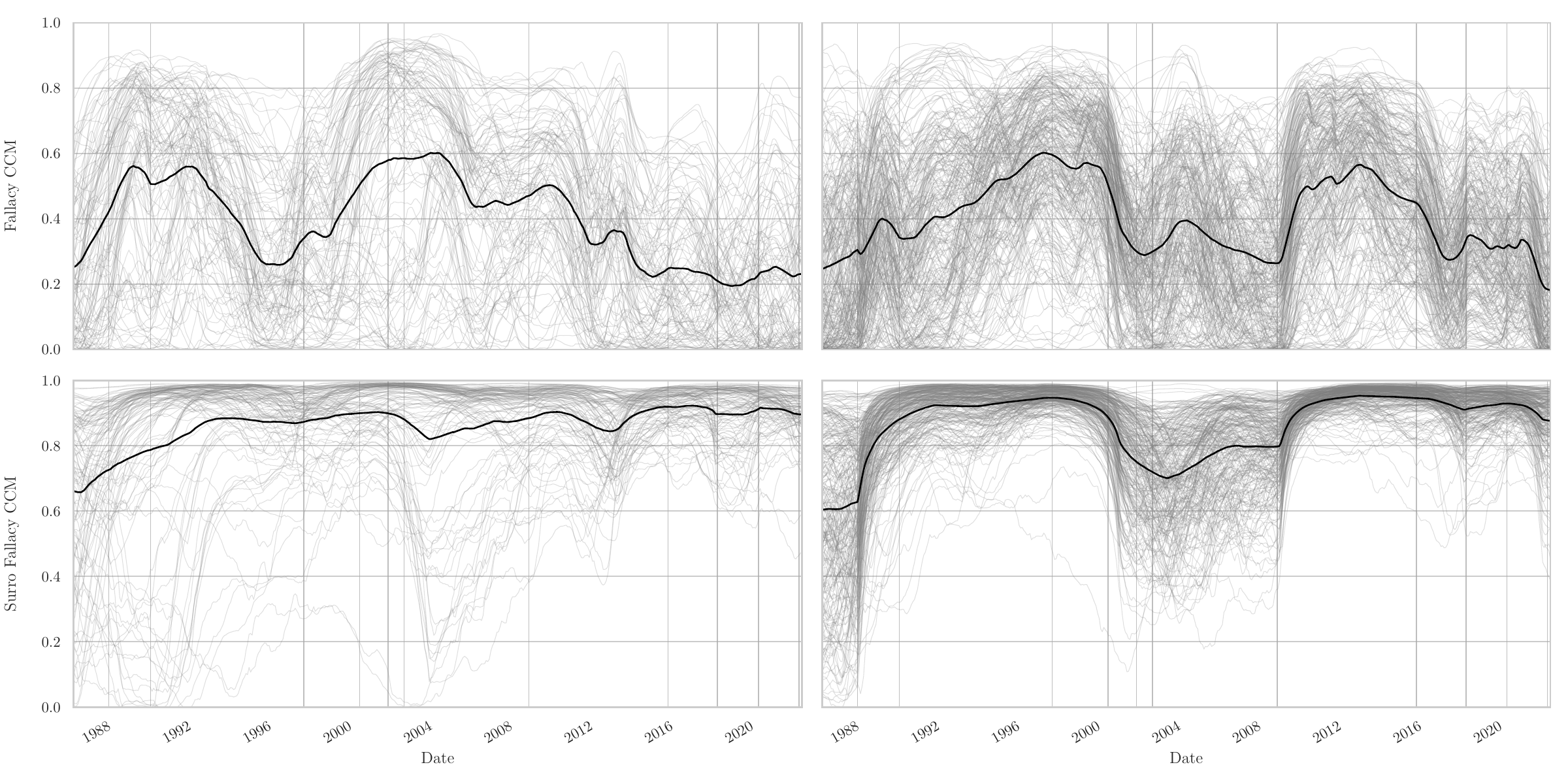}
    \caption{Fallacy Convergent Cross Mapping. The setup of this figure is analogous to Fig. \ref{fig:transfer_entropy_fallacy}.
    }
    \label{fig:convergent_cross_mapping_fallacy}
\end{figure*}


\subsection{\label{sec:decomposition}Linear and Nonlinear Decomposition}
\noindent To decompose the causal relationships within time series systems into components originating from linear and nonlinear drivers, we employ surrogate techniques based on the \textit{Fourier Transform} (FT). Employing these surrogates on (causality) measures, we devise methodologies to systematically capture the quantitative breakdown of linear and nonlinear influences.

\subsubsection{Fourier Transform Surrogates}
\noindent FT surrogates destroy the nonlinear characteristics of a time series $\boldsymbol{x}$ while keeping the linear ones unaffected \cite{raeth2009surrogates}. The algorithm to generate FT surrogates is described as follows \cite{rath2012revisiting}:

\begin{enumerate}
    \item \textbf{Fourier Transform}: Given a real-valued time series $\mathbf{x} = \{x_1, x_2, \ldots, x_N\}$, compute its Fourier transform $\mathbf{F}(\mathbf{x})$ using the \textit{Fast Fourier Transform} (FFT) algorithm \cite{brigham1988fast}. 
   \[
   \mathbf{F}(\mathbf{x}) = \textit{FFT}(\mathbf{x})
   \]
    \item \textbf{Phase Randomization}: Preserve the amplitudes but randomize the phases of the Fourier coefficients. This can be done by multiplying the complex Fourier coefficients by a random phase factor $e^{i\phi}$, where $\phi$ is uniformly distributed over the interval $[0, 2\pi]$. The phase-randomized Fourier Transform $\mathbf{F}'(\mathbf{x})$ is given by:

   \[
   F'_k = |F_k| \cdot e^{i\phi_k}, \quad \phi_k \in [0, 2\pi]
   \]
   \item \textbf{Inverse Fourier Transform}: Compute the inverse FT of the phase-randomized coefficients to obtain the surrogate time series $\mathbf{\Tilde{x}}$:

   \[
   \mathbf{\Tilde{x}} = \textit{IFFT}(\mathbf{F}'(\mathbf{x}))
   \]
    By keeping the amplitudes of the original data and only randomizing the phases, the resulting surrogates maintain the power spectral density of the original time series but break the higher-order statistical dependencies. 
\end{enumerate}

\noindent To enhance the reliability of our findings, we average metrics derived from surrogate time series over various instances $K$ of random phases. The surrogate of time series $\mathbf{x}$, when subjected to the random phases of realization $k$, is denoted as $\mathbf{\Tilde{x}}^{(k)}$.

\begin{figure*}[!t]
	\includegraphics[width=0.95\textwidth, keepaspectratio]{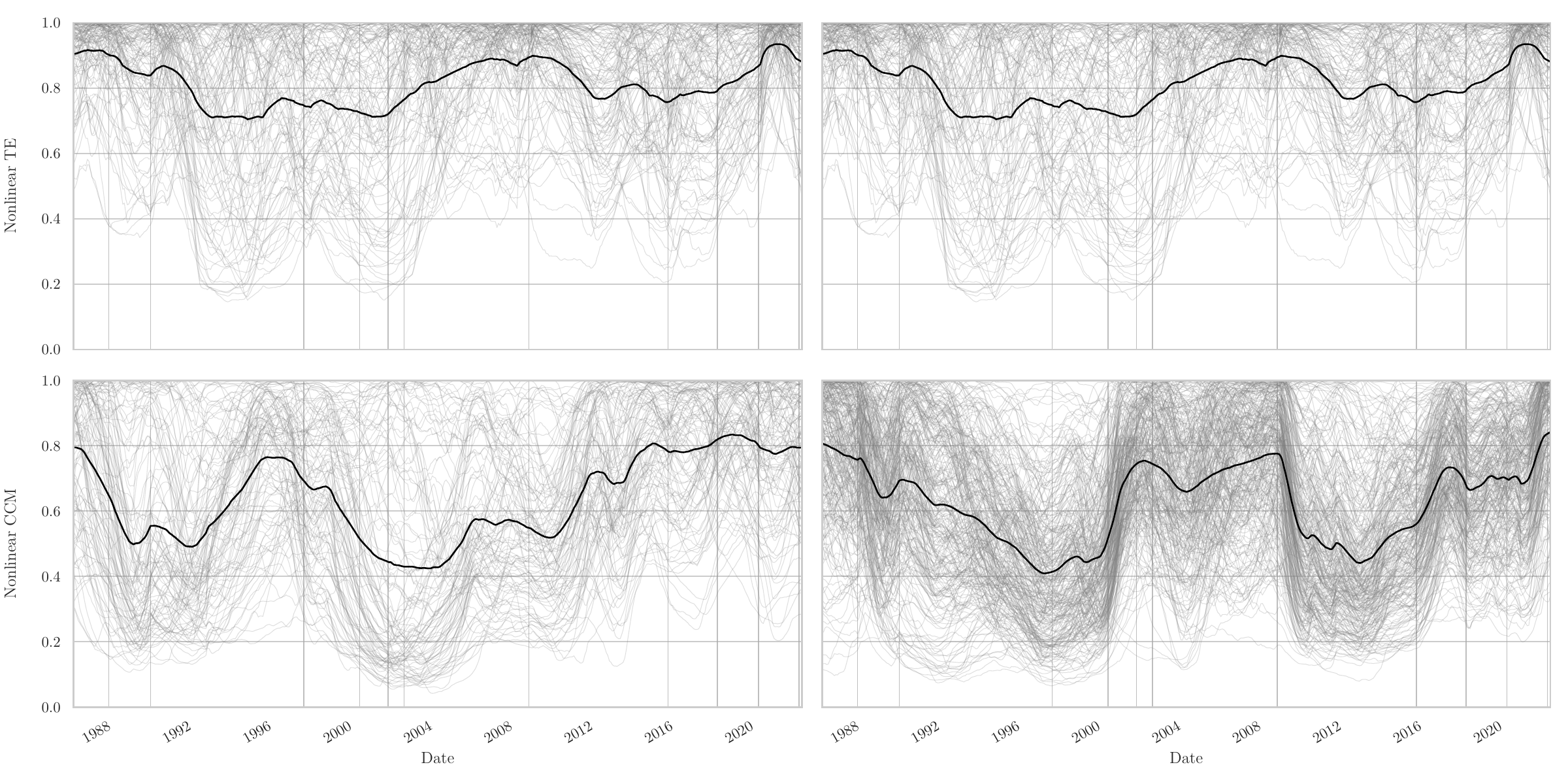}
    \caption{Nonlinear causality. The first row shows the historical nonlinear TE of stocks within the German DAX (left) and the U.S. Dow-Jones (right) indices, respectively. Each line represents one direction of the TE between two stocks over time. The bottom row illustrates the nonlinear CCM. The vertical lines represent important economic or political events.
    }
    \label{fig:nonlinear_causality}
\end{figure*}


\subsubsection{Linear and Nonlinear Measures}
\noindent In order to evaluate how much of a (causal) measure is attributed to linear or nonlinear effects, we adopt a specific approach that involves the calculation of measures on surrogate time series. Within the context of this research, we focus on a bivariate measure, denoted as $\psi\big(\mathbf{x}, \mathbf{y}\big)$, which is a function mapping two time series to a real number. This function's purpose is to capture the relationship between the two time series in numerical terms. The corresponding surrogate or linear measure is defined as the average over $K$ surrogate realizations of both time series:

\begin{equation}
        \Tilde{\psi}(\mathbf{x}, \mathbf{y}) \equiv \frac{1}{K} \sum_{k=1}^{K} \psi\big(\tilde{\mathbf{x}}^{(k)}, \tilde{\mathbf{y}}^{(k)}\big)\, .
\end{equation}

\noindent Here, the superscript $k$ indicates that we add the same random phases to both time series within a single realization. This choice ensures that phase differences remain unaffected, preserving specific properties such as the Pearson correlation \cite{prichard1994generating}. To ensure robustness we repeat the calculation for $K=50$ surrogate realizations.

\subsubsection{Nested Measures}
\noindent As aforementioned, employing rolling windows transforms the measure $\psi$ into a vector. This transition allows for the investigation of interrelations between two measures through a third expression:

\begin{equation}
        \psi_{ir} \equiv \rho (\mathbf{\psi}_1, \mathbf{\psi}_2) \, .
\end{equation}
\noindent Particularly, we can utilize the Pearson correlation $\rho$ to study the relationship between the original measure and its corresponding surrogate, expressed as:

\begin{equation}
    \rho (\mathbf{\psi}, \mathbf{\Tilde{\psi}}) \, .
\end{equation}

\noindent  This method also allows for expressing the coefficient of determination using the Pearson correlation, as mentioned in \cite{kasuya2019use}:

\begin{equation}
    R^2=\rho^2 \in \left[0, 1\right] \, ,
\end{equation}

\noindent This enables us to quantify the extent of the measure attributable to linear influences, more precisely, the fraction of the variability in the measure $\psi$ that can be explained from the surrogate measure $\Tilde{\psi}$. What remains then emanates from nonlinear characteristics:

\begin{equation}
   \psi_{nl} \equiv  1 - \rho^2 (\mathbf{\psi}, \mathbf{\Tilde{\psi}}) \, .
\end{equation}

\noindent Furthermore, there's an application to the exploration of the correlation-causality fallacy \cite{maziarz2015review}. This involves determining how much of the causality is explained by correlation:

\begin{equation}
     \psi_{fall} \equiv \rho^2 (\mathbf{\psi}, \mathbf{\rho}) \, ,
\end{equation}

\noindent serving as a gauge of the causal relationship that can be explained by correlation. Specifically, this measure for the fallacy can be applied to the surrogate measure in order to evaluate how much of the linear causality is captured by correlation: 

\begin{equation}
     \psi_{fall, lin} \equiv \rho^2 (\Tilde{\mathbf{\psi}}, \mathbf{\rho}) \, .
\end{equation}

\begin{figure*}[!t]
	\includegraphics[width=0.95\textwidth, keepaspectratio]{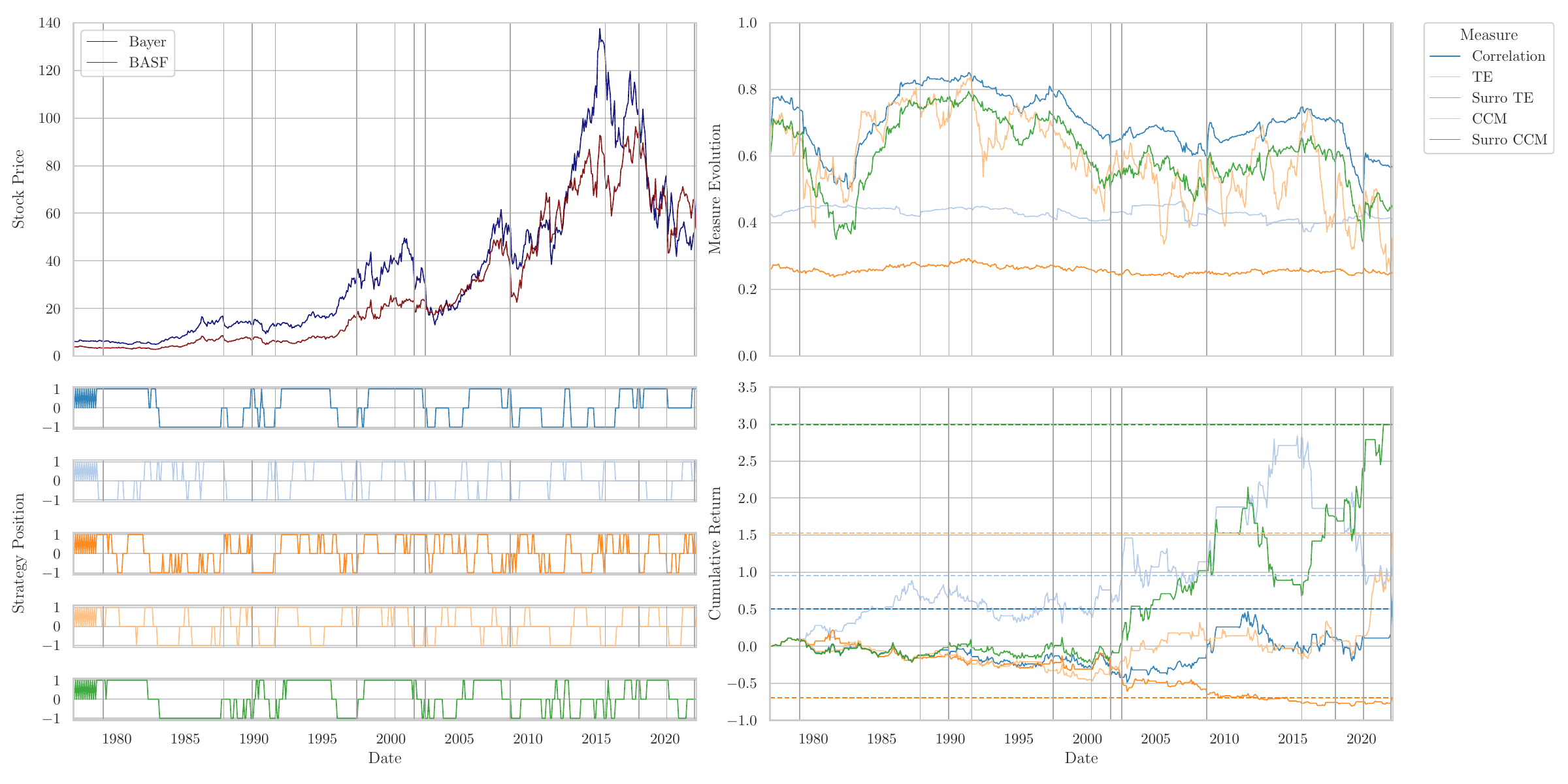}
    \caption{Pair Trading. The stock prices of two companies from the DAX (Bayer and BASF) are displayed in the top left figure. The top right figure presents the co-dependence measures over time, with each color corresponding to a specific co-dependence measure that is included in the legend on the right-hand side. The bottom left chart illustrates the strategy positions over time, with long position in Bayer and short position in BASF indicated by $1$, the opposite indicated by $-1$, and no investment indicated by $0$. The graph in the lower right corner illustrates the cumulative return achieved by the strategy over time. The dotted horizontal lines mark the strategy's most recent cumulative return value. The vertical lines indicate notable economic or political events.
    }
    \label{fig:pair_trading}
\end{figure*}


\subsection{\label{sec:financial}Financial Frameworks}
\noindent Here, we introduce two financial frameworks and demonstrate how causality can be easily integrated, while simultaneously enhancing performance.

\subsubsection{Pair Trading}
\noindent Pair trading is a popular and widely utilized strategy in quantitative finance that aims to capitalize on relative price movements between two closely related assets \cite{vidyamurthy2004pairs}. This strategy is grounded in the concept of mean reversion, which assumes that over time, the prices of assets that are historically correlated tend to revert to their historical average relationship. The basic premise is to find two stocks that are highly correlated. When they deviate from this correlation (i.e., one stock moves up while the other moves down or vice versa), we take a \textit{long} position in the underperforming stock and a \textit{short} position in the outperforming stock, expecting them to revert to their historical correlation \cite{hull2019options}. Thus, a basic form of the strategy involves the following steps:

\begin{enumerate}
    \item \textbf{Correlation Calculation}: We calculate the rolling historical and the short-term correlation between two stocks 
    \item \textbf{Signal Generation}: When the current correlation $\rho_t$ deviates from its historical mean by a certain threshold, a trading signal is generated. A common approach is to use the \textit{z-score} $z$ of the spread, which measures the number of standard deviations by which the current correlation deviates from its historical mean:
    
    \begin{equation}
        z_t = \frac{\rho_t - \bar{\rho}_{hist}}{\sigma_{\rho_{hist}}} \, ,
    \end{equation}

    \noindent where $\bar{\rho}_{hist}$ and $\sigma_{\rho_{hist}}$ denote the mean and standard deviation of the historical correlation, respectively.
    
    \item \textbf{Trade Execution}: When the z-score crosses a predefined threshold (e.g., above a positive threshold for a long trade or below a negative threshold for a short trade), a trade is initiated. A long trade involves buying the underperforming asset and simultaneously shorting the overperforming asset. We set the threshold at $z_t \pm 1.5$.
    
    \item \textbf{Profit Taking}: The strategy aims to profit from the mean reversion process. As the spread narrows and returns to its historical mean, the positions are unwound, resulting in a profit.
\end{enumerate}

\noindent We would like to note that we are aware of the simplifications of the strategy and that for practical use more fine-tuning is necessary. However, we find the parametrization of the strategy to be sufficient for illustrative purposes. For our purposes we exchange the historical Pearson correlation with the TE and CCM respectively.

\begin{figure*}[!t]
	\includegraphics[width=0.95\textwidth, keepaspectratio]{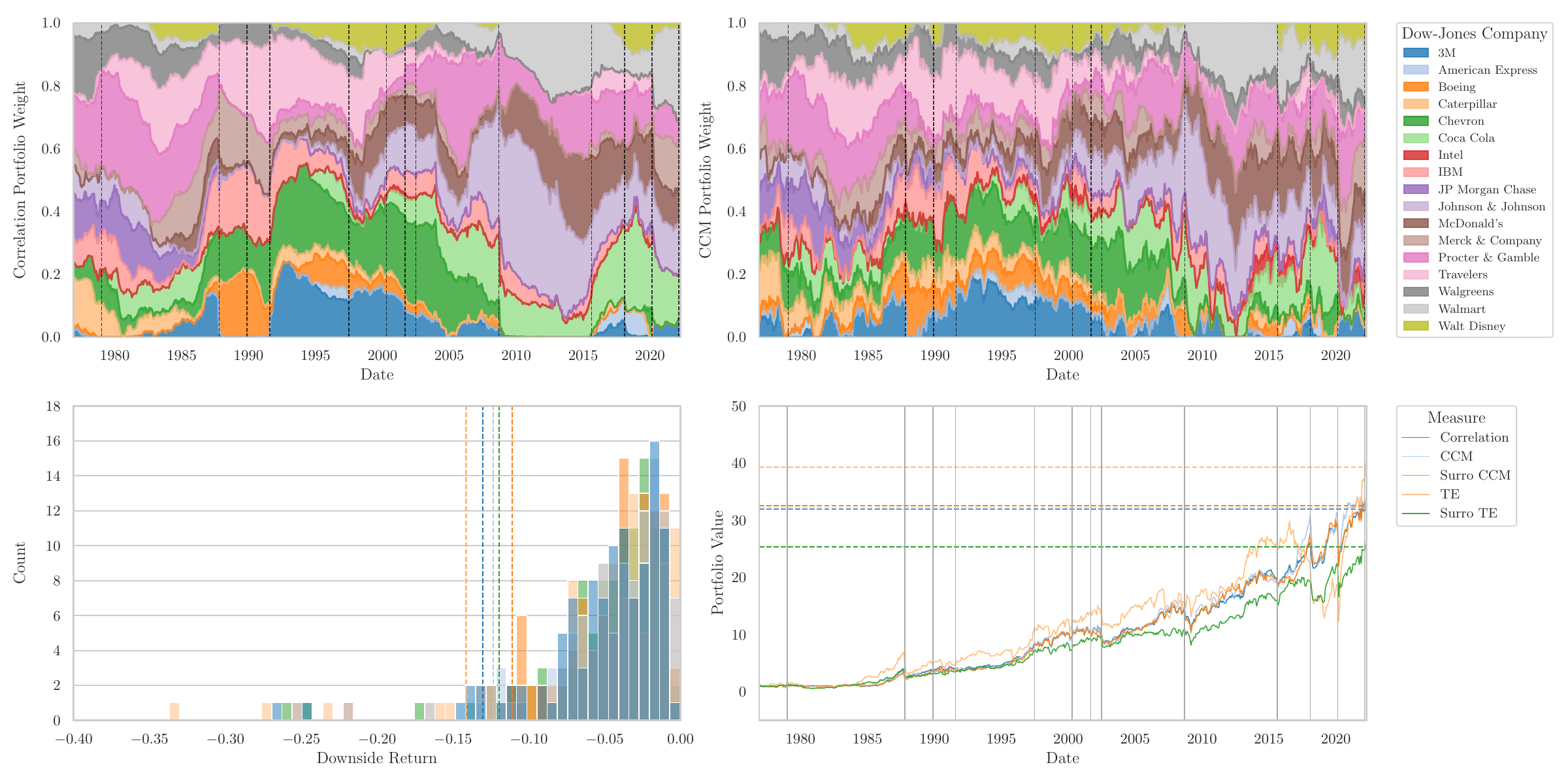}
    \caption{Minimum Risk Portfolio Optimization. The top row displays the optimized Minimum Risk Portfolio weights over time using both the correlation (on the left) and CCM (on the right) as co-dependence measures. Each colored area represents a stock from the Dow-Jones, which is mapped in the legend to the right. The dotted vertical lines depict significant economic or political events. In the bottom row, the left figure illustrates the distributions of the downside returns when using different co-dependence measures. The vertical lines depict the VaR at $\alpha=1\%$ level. The plot to the right displays the portfolio's value over time. The vertical lines denote significant economic or political occurrences. The dotted horizontal lines denote the portfolio's most recent value. Each color corresponds to a particular codependence measure, which is mapped in the right-hand side legend.
    }
    \label{fig:minimum_risk}
\end{figure*}


\subsubsection{Portfolio Optimization}
\noindent In the world of finance, \textit{Markowitz Portfolio Theory} (MPT), developed by Harry Markowitz in 1952, is a cornerstone concept for investors and financial analysts \cite{markowitz1991foundations}. This theory revolutionized the way investors think about constructing portfolios. It is based on a fundamental premise: rational investors seek to maximize their portfolio's expected return while minimizing its risk. The key insight here is that an asset's risk and return should not be evaluated in isolation but rather in the context of the entire portfolio. 

The expected return of a portfolio is calculated as a weighted sum of the expected returns of its individual assets:

\begin{equation}
    E(R_p) = \sum_{i=1}^{n} w_i \cdot E(R_i) \, ,
\end{equation}

\noindent where $E(R_p)$ is the expected return of the portfolio, $w_i$ is the weight of asset $i$ in the portfolio, and $E(R_i)$ is the expected return of asset $i$. Even though historic returns do not indicate future performance, it is common to use the historical mean as a proxy for the expected returns \cite{hull2019options}. 

The portfolio's variance is a measure of its risk. It considers not only the individual asset variances but also the correlation between assets. The formula for portfolio variance is:

\begin{equation}
    \sigma^2_p = \sum_{i=1}^{n} \sum_{j=1}^{n} w_i \cdot w_j \cdot\sigma_i \cdot \sigma_j  \cdot \rho_{ij} \, ,
\end{equation}

\noindent where $\sigma^2_p$ is the variance of the portfolio, $w_i$ and $w_j$ are the weights of assets $i$ and $j$ in the portfolio, and $\sigma_{ij}$ is the covariance between assets $i$ and $j$. We can replace the correlation with a causality measure $\psi$ or use the sign of the correlation if the measure $\psi$ is normalized to $\left[0, 1\right]$:

\begin{equation}
    \sigma^2_p = \sum_{i=1}^{n} \sum_{j=1}^{n} w_i \cdot w_j \cdot\sigma_i \cdot \sigma_j  \cdot \psi_{ij} \cdot \textit{sgn}\left(\rho_{ij}\right) \, ,
\end{equation}

\noindent where $\textit{sgn}(\cdot)$ denotes the Sign function.

A popular measure of the riskiness of historical portfolio performance is \textit{Value-at-Risk} (VaR), which quantifies the potential loss in value of an investment or portfolio over a specified time horizon at a $\alpha$ \cite{duffie1997overview} confidence level. A $1-\alpha$ $VaR = x$ means that there is a $\alpha$ chance that the portfolio will lose more than $x$. Unlike standard deviation, VaR measures tail risk and does not assume a normal distribution, which is particularly important for risk management purposes. We use the default value of $\alpha=1\%$.

Two portfolios of great importance within MPT are the \textit{Minimum Risk Portfolio} and the\textit{Maximum Sharpe Ratio Portfolio}. These portfolios play a crucial role in portfolio analysis and optimization: 

\begin{itemize}
    \item \textit{Minimum Risk}: The Minimum Risk Portfolio represents the portfolio with the lowest possible risk for a given set of assets. Mathematically, it can be formulated as an optimization problem. The solution to this problem provides the weights of assets in the Minimum Risk Portfolio:

    \begin{align*}
    \text{Minimize} & \quad \sigma^2_p \\
    \text{Subject to} & \quad E(R_p) = \text{target return} \\
    & \quad \sum_{i=1}^{n} w_i = 1 \\
    & \quad w_i \geq 0 \quad \text{for all } i
    \end{align*}
    
    \item \textit{Maximum Sharpe Ratio}: The Maximum Sharpe Ratio Portfolio represents the portfolio that offers the highest risk-adjusted return. The Sharpe Ratio ($S$) measures this risk-adjusted performance:

    \begin{equation}
    S = \frac{E(R_p - R_f)}{\sigma_p}    
    \end{equation}  

    \noindent To find the Maximum Sharpe Ratio Portfolio, we maximize the Sharpe Ratio by adjusting the asset weights. Mathematically:

    \begin{align*}
    \text{Maximize} & \quad S \\
    \text{Subject to} & \quad \sum_{i=1}^{n} w_i = 1 \\
    & \quad w_i \geq 0 \quad \text{for all } i
    \end{align*}

\end{itemize}

\noindent We illustrate a simple way to incorporate causality measures into portfolio construction through the utilization of these two portfolios. As a result, we regularly adjust the portfolio by optimizing its weightings with the mentioned algorithms to align it with the prevailing market conditions. To achieve this, we apply the rolling causality measures as previously demonstrated in this paper. Consequently, we can assess the advantages of using causality measures as the co-dependency metric for the portfolio, examining both its performance and risk management implications.

\begin{figure*}[!t]
	\includegraphics[width=0.95\textwidth, keepaspectratio]{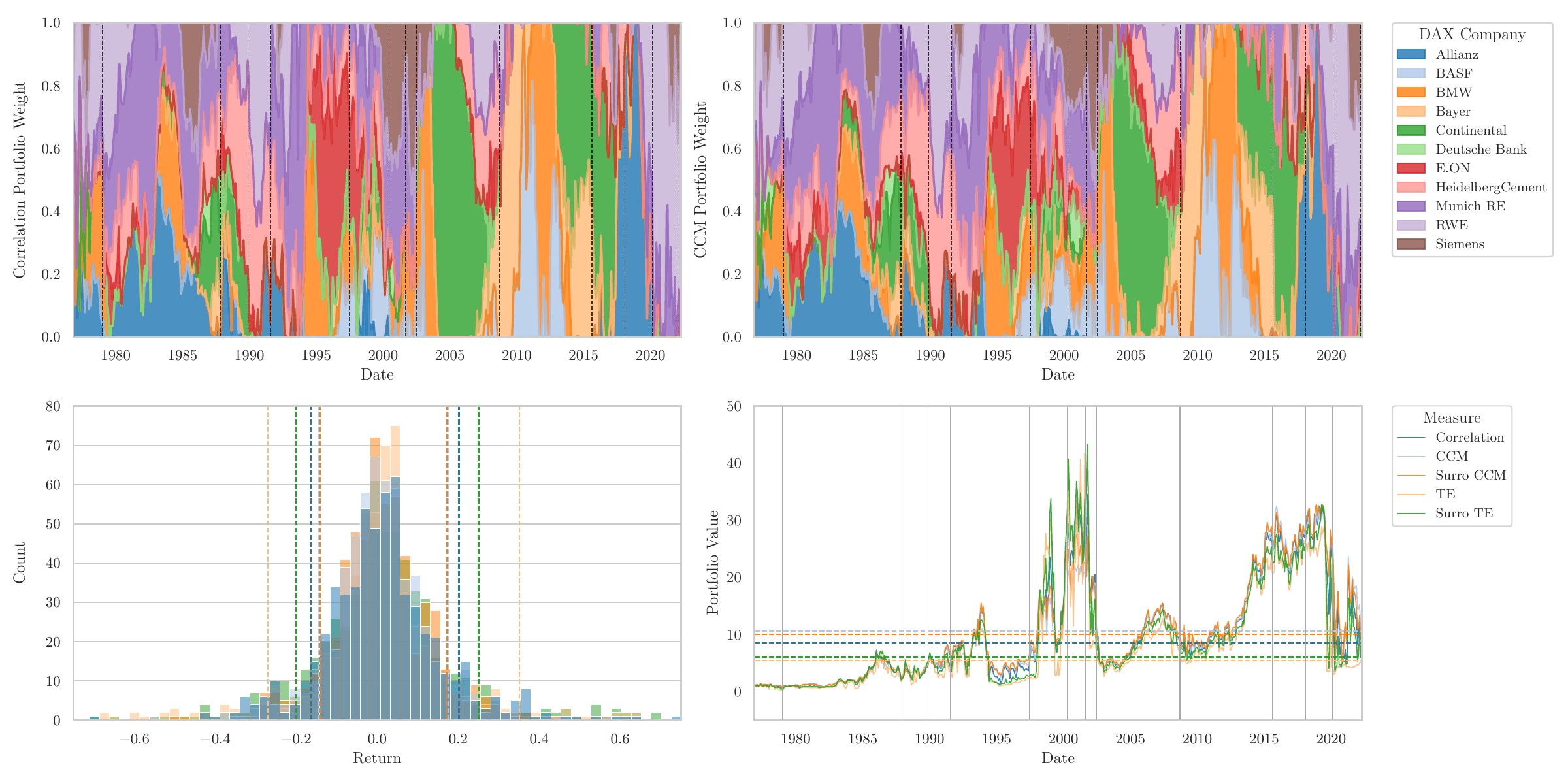}
    \caption{Maximum Sharpe Ratio Portfolio Optimization. The setup of this figure is analogous to Fig. \ref{fig:minimum_risk}. The top row displays the optimized Maximum Sharpe Ratio Portfolio weights over time using both the correlation (on the left) and CCM (on the right) as co-dependence measures. Each colored area represents a stock from the DAX, which is mapped in the legend to the right. The dotted vertical lines depict significant economic or political events. In the bottom row, the left figure illustrates the distributions of the returns when using different co-dependence measures. The vertical lines depict the standard deviations of the returns. The plot to the right displays the portfolio's value over time. The vertical lines denote significant economic or political occurrences. The dotted horizontal lines denote the portfolio's most recent value. Each color corresponds to a particular co-dependence measure, which is mapped in the right-hand side legend.
    }
    \label{fig:maximum_sharpe}
\end{figure*}


\section{Results}
\noindent In the following we present the results of our analyses, which we structure into three subsections. As motivated by Figure \ref{fig:coupled_difference}, we observe that for complex and chaotic systems it is difficult to measure the co-dependence of variables through correlations as they can exhibit different regimes of positive, negative, and no correlation even though they are guided by exactly the same governing equations. This is illustrated by the rolling window analysis of the correlation, which is unrobust and changes significantly over time. Hence in order to measure their co-dependence reliably, another measure is needed. Causality measures, such as CCM, are a valuable technique to measure the causality of two variables in both directions and provide stable results over time. Furthermore, by using FT surrogates, we can separate the causality in linear and nonlinear contributions which helps to understand the intricate nature of the co-dependence. The Figure shows that the separation of causality is stable over different windows and also plausible when compared to the governing Equations \ref{eq:coupled_difference}. 

\subsection{Historical Causality}
\noindent To demonstrate the practical applicability of our framework, we have employed it in an analysis of major German and U.S. stock indices. The data and the dynamic correlation patterns are visually depicted in Figure \ref{fig:data_and_correlation}. Notably, these correlations undergo significant shifts during and after pivotal economic and political events. This phenomenon can be attributed to the changing behavior of investors and other market participants in response to these impactful occurrences. Furthermore, this effect extends to our investigation of causality measures, as demonstrated in Figures \ref{fig:transfer_entropy} and \ref{eq:convergent_cross_mapping}. These figures reveal that linear and nonlinear causality measures, such as Transfer Entropy (TE) and Convergent Cross Mapping (CCM), exhibit analogous responses to these events.

Specifically, when examining TE, it becomes apparent that TE is highly responsive to these events, displaying sharp fluctuations. In contrast, surrogate TE remains relatively stable and does not react as drastically. Conversely, surrogate CCM appears to respond more strongly than regular CCM, displaying significant jumps similar to the observed patterns in correlation. One of the most striking examples of this behavior is observed during Black Monday in 1987, where we witness substantial increases in correlation, TE, and surrogate CCM, particularly in the context of U.S. stocks. Two other significant events that exhibit similar patterns are the global financial crisis in 2009 and the COVID-19 pandemic in 2020. These observations suggest that these events triggered structural shifts in the market, which is reasonable given their profound impacts on the global economy. An intriguing observation is that TE experiences more pronounced fluctuations compared to surrogate TE during these events, while the opposite is observed for CCM. This suggests that the linear dynamics in the stock markets were more profoundly influenced, possibly due to investors simultaneously adjusting their stock positions in response to the market crashes.

\subsection{Correlation-Causality Fallacy and Nonlinear Causality}
\noindent Upon examination of Figure \ref{fig:transfer_entropy_fallacy}, it becomes evident that both the original and surrogate Transfer Entropy (TE) exhibit a moderate correlation. Notably, there is an intriguing exception during the period spanning from approximately 1990 to 2002 in the U.S. stock market, where a substantial portion, approximately $75\%$, of TE can be attributed to correlation. This spike coincided with the rise and eventual burst of the dotcom bubble, suggesting that it might have served as an indicator of abnormal market behavior during this period.

One of the most significant findings from this analysis is the observation that fallacy of surrogate Convergent Cross Mapping (CCM) is remarkably high, around $90\%$, in both the German and U.S. stock indices, as depicted in Figure \ref{fig:convergent_cross_mapping_fallacy}. This suggests that correlation effectively acted as a suitable proxy for linear causality for the majority of the past few decades. However, in periods where this fallacy diminishes, such as the aftermath of the dotcom bubble in 2002 and the onset of the global financial crisis in 2008, relying solely on correlation as a measure of co-dependence significantly underestimates portfolio risk, as nonlinear effects cannot be disregarded. This effect is even more pronounced when examining the fallacy of the original CCM, where we also observe a substantial drop during these phases.

To gauge the extent of nonlinear contributions to our causality measures, we delve into the analysis of how much of the causality can be accounted for by its surrogate. In Figure \ref{fig:nonlinear_causality}, we observe the evolution of nonlinear causality over time, noting that nonlinear TE and CCM exhibit similar but not identical behaviors. Both measures reveal heightened levels of nonlinearity during the period between the dotcom bubble burst and the commencement of the global financial crisis. In contrast, before and after this period, we observe phases with less nonlinearity. This indicates that these two major economic events should be assessed differently, as the dotcom bubble led to increased nonlinearity in its aftermath, while the global financial crisis, precipitated by the U.S. housing market crisis, ushered in a phase of more linear market behavior. Particularly for CCM, this behavior is quite drastic, with jumps exceeding $20\%$. In conclusion, our analysis suggests that nonlinear causality can be a valuable tool for anticipating and evaluating financial impacts, provided it is continually monitored and assessed in the context of evolving market dynamics.

\subsection{Pair Trading and Portfolio Optimization}
\noindent To effectively apply causality measures in practical financial scenarios, we present two common financial frameworks where the interdependence between assets plays a pivotal role. The first concept we explore is pair trading, a logical choice given its reliance on the idea that two assets tend to revert to a default correlation, and deviations from this norm can be profitably exploited. In Figure \ref{fig:pair_trading}, we use two German stocks from the chemical industry, Bayer and BASF, to illustrate how causality measures can be seamlessly integrated. It's noteworthy that even though the differences in the evolution of co-dependence measures are relatively similar, over time, these subtle distinctions significantly impact trading performance. Of particular interest is the fact that the trading strategy employing surrogate Convergent Cross Mapping (CCM) outperforms the one utilizing correlation by a substantial margin, approximately six times, despite the measures' apparent similarity. Additionally, we observe that Transfer Entropy (TE) and CCM perform better than correlation, while surrogate TE lags behind and even delivers negative returns. This straightforward example underscores the potential of a causality-based pair trading strategy.

As previously highlighted, relying solely on correlation can potentially lead to an underestimation of risk, a perilous scenario when managing a portfolio. In Figure \ref{fig:minimum_risk}, we employ stocks from the U.S. Dow-Jones index and minimize risk by dynamically optimizing the portfolio weights on a monthly basis. It becomes evident that the allocations of a portfolio using correlation and CCM exhibit visible disparities over time. This divergence is reflected in the portfolio's downside returns and overall performance. Notably, we observe that a portfolio employing surrogate Transfer Entropy (TE), CCM, and surrogate CCM achieves a superior $1\%$ Value at Risk (VaR) while slightly enhancing portfolio performance.

Similarly, in the context of optimizing the Sharpe Ratio, as depicted in Figure \ref{fig:maximum_sharpe}, the inclusion of causality measures results in a more favorable risk-return profile. When optimizing the stocks of the German DAX index, we note a reduction in portfolio standard deviation and an increase in portfolio value over time, particularly when employing original and surrogate CCM.

\section{Conclusion and Outlook}
The present study has addressed the issue of identifying and quantifying co-dependence among financial instruments, which continues to be a paramount challenge for both researchers and practitioners in the financial industry. While traditional linear measures like the Pearson correlation have maintained their prominence, this paper has introduced a novel framework aimed at analyzing both linear and nonlinear causal relationships within financial markets. To achieve this, we have employed two distinct causal inference methodologies, namely Transfer Entropy and Convergent Cross-Mapping, and have utilized Fourier transform surrogates to disentangle their respective linear and nonlinear contributions.

Our findings have unveiled that stock indices in Germany and the U.S. exhibit a substantial degree of nonlinear causality, a phenomenon that has largely eluded previous investigations. It is important to recognize that while correlation, exemplified by the Pearson correlation coefficient, serves as an excellent proxy for linear causality, it falls short in capturing the intricate nonlinear dynamics that underlie financial markets. Consequently, relying solely on correlation can lead to an underestimation of causality itself.

The framework introduced in this study not only facilitates the quantification of nonlinear causality but also sheds light on the perilous "correlation-causality fallacy." By delving into the nuances of causality, we have motivated how these insights can be harnessed for practical applications, including inferring market signals, implementing pair trading strategies, and enhancing the management of portfolio risk. 

One of the insights derived from our findings underscores the role that both linear and nonlinear causality can play as early warning indicators for unusual market dynamics. Furthermore, our results suggest that a straightforward incorporation of these causality measures into strategies, such as pair trading and portfolio optimization, can yield better outcomes compared to a reliance solely on Pearson correlation. This understanding can significantly empower traders and risk managers, enabling them to craft more effective trading strategies and to adopt a more proactive approach to risk mitigation.

Looking ahead, the implications of our findings extend to various facets of financial research and practice. Further exploration of nonlinear causality may uncover new dimensions of financial market interactions, potentially leading to the development of innovative trading algorithms and risk management tools. Additionally, the integration of causality measures into existing financial models and frameworks holds the promise of enhancing their predictive accuracy and robustness.

In conclusion, this paper has introduced a comprehensive framework for disentangling linear and nonlinear causality within financial markets. The revelation of substantial nonlinear causality and the recognition of the limitations of traditional correlation measures underline the importance of taking a more nuanced approach to co-dependency analysis. The insights gained from this study have the potential to enhance the way we perceive and navigate the intricacies of financial markets, contributing to more informed decision-making, better risk management practices, and more financial stability.

\begin{acknowledgments}
\noindent We would like to thank the DLR and AllianzGI for providing data and computational resources.
\end{acknowledgments}

\nocite{*}
\bibliography{aipsamp} 

\end{document}